\documentclass[12pt]{article}
\usepackage{eurosym}
\usepackage[left=0.5in, right=0.5in, top=0.75in, bottom=0.75in]{geometry}
\usepackage{hyperref}
\usepackage{mathtools}
\usepackage{graphicx}
\usepackage{mathrsfs}
\usepackage{mathtools}

\makeatletter
\newcommand{\row}[1]{\mathord{\buildrel{\lower3pt\hbox{$\scriptscriptstyle\rightarrow$}}\over #1}}

\newcommand{\dyadic}[1]{\mathord{\dyadic@rrow{#1}}}
\newcommand{\dyadic@rrow}[1]{
\begin{picture}(12,12)(-1,0)
\put(-2,12){\makebox(0,0)[t]{$\scriptscriptstyle\downarrow$}}
\put(-2,12){\makebox(0,0)[l]{$\scriptscriptstyle\longrightarrow$}}
\put(5,0){\makebox(0,0)[b]{$#1$}}
\end{picture}
}
\newcommand{\bra}[1]{\bigl\langle #1 \bigr|}
\newcommand{\ket}[1]{\bigl| #1 \bigr\rangle}
\begin{document}

	\renewcommand{\baselinestretch}{1.3} \topmargin=-1.8cm \textheight=23 cm
	\textwidth=23cm
	
	\begin{center}
		          \textit{\LARGE Fisher information of accelerated   two-qubit system  in the presence of the color and white noise channels}
		
	\textit{F. Ebrahim}$^{a}$ ,
	\textit{N. Metwally}$^{a,b}$
\textit{\footnote{nmetwally@nmetwally@gmail.com}}\\
				 $^{a}${\footnotesize Math. Dept., College of Science, University of Bahrain, Bahrain.}\\
				 $^{b}${\footnotesize Department of Mathematics, Aswan University
		 	Aswan, Sahari 81528, Egypt}
	
	\end{center}

\begin{abstract}
In this manuscript, we investigate the effect of the white and color noise on a accelerated two- qubit system, where different initial state setting are considered. The behavior of the survival  amount of entanglement is quantified for this accelerated system by means of the concurrence.
We show that, the  color noise enhances the generated entanglement between the two particles even for small values of the initial purity of the accelerated state. However, the larger values of the white noise strength improve the generated entanglement.
 The initial parameters that describe this system  are estimated by using Fisher information, where two forms are considered, namely  by using a single and two-qubit forms.
 It  is shown that,  by using the two-qubit form, the estimation degree of these parameters is larger than that displayed by using a single-qubit form

\end{abstract}
\section{Introduction.}

It is well known that, decoheronce is unavoidable phenomena of quantum systems \cite{Yu2004,Yu2007,Yu2009}. It may appear during  the  generation  process of the entangled systems or their transmission it to the users to perform some quantum information tasks, where it interacts with its surroundings\cite{Palm} .
 There are many studies  which are devoted to investigate the behavior of the  entangled systems in the presence of noise channels. As an example, Prakash et. al. investigated the effect of decoherence on the fidelity of the teleported states by  using entangled coherent states \cite{Prakash} .
 The decay of the quantum correlations may appear due to the dissipation medium
which effect on the lifetimes of the states of the system \cite{Eleuch}. The  information loss in local dissipation environments is discussed by Metwally \cite{Metwally2010}. Dynamics of encrypted information in superconducting
qubits with the presence of imperfect operations is discussed in\cite{Metwally2012}.

Recently, quantifying the  of entanglement of different sizes of accelerated system has paid many attention,
where it has shown that, entanglement between the accelerated entangled
observers  is degraded \cite{Als}.  The sudden death of entanglement and the
dynamics of mutual information have been investigated by Landulfo
and Matsas \cite{andr} in non-inertial frames. Metwally, \cite{Metwally2013} investigated the usefulness  accelerated classes to perform quantum information tasks.  The possibility of using the accelerated systems to preform quantum teleportation is discussed in \cite{Metwally2013-1}. Quantum coding in non-inertial frame is investigated by Metwally et. al \cite{Metwally2014,Metwally2015}.

Additionally,
quantum Fisher information(QFI) is used as an estimator of parameters that
are contained in a quantum system during its evolution \cite{Nilsen,MA}. It  plays a significant role in the fields
of quantum information theory  and quantum metrology\cite{Lloyd}. Due to its importance, there are some efforts that have been
taken to quantify QFI for  different quantum systems. For example, the  dynamics of QFI
of a two-qubit system, where each qubit interacts with its own Markovian environment is discussed in \cite{Zheng}.
Ozaydin \cite{Ozaydin}  quantified the QFI analytically for the W-state in the presence of different
noisy channels.  Metwally \cite{Metwally2017}, discussed the estimation of teleported and gained parameters in
a non-inertial frame.

Therefore, in this contribution we are motivate to  investigate the effect of the color and white noises on the  degree of  entanglement of the accelerated system. Also, the  estimation degree of the initial parameters of  the initial state settings  are quantified by using the Fisher information, where two forms are considered; one based on the definition of Fisher information of single qubit and the second based on that for the two-qubits form.

 The paper is arranged as. In Sec.2 , we describe the suggested initial system in the presences of the white and color noises. The effect of the channels strengths on the degree of entanglement is discussed in Sec.3, where we introduce the  concurrence as a measure  of the entanglemet.  Analytical forms of the concurrence are given for different noise channels . In Sec.4, we review two forms of Fisher information, one for a single qubit and the second for two-qubit system. Moreover, we discuss the behavior of the Fisher information numerically to display the effect of the noisy channels on the estimation degree of the initial state settings. Finally, we summarize our results in Sec.5.

\section{The suggested model}

Let us consider a system consists of two qubits is initially prepared in a maximum, partial or separable state. This  system is forced to pass through either white  or color noises \cite{Adn,Dot}.  In the computational basis, this system may be  described  as

\begin{equation}\label{ini}
\rho_{w(c)}=\left\{ \begin{array}{ll}
p\ket{\phi}\bra{\phi}+\frac{1-p}{4}\hat{I},&\mbox{for white noise}\\
\nonumber\\
q\ket{\phi}\bra{\phi}+\frac{1-q}{2}(\ket{01}+\ket{10}),&\mbox{for color noise},
\end{array} \right.
\nonumber\\
\end{equation}
where $\ket\phi=\sqrt{1-x^2}\ket{01}+x\ket{10}$, and the parameters $p$ and $q$ represent the strengths of the white and the color noisy, respectively. This state behaves as a product state at $x=0$ or $x=1$ and as a singlet state at $x=\frac{1}{\sqrt{2}}$, namely  maximum entangled state . However, for any value of $0<x<1$, the state $\ket{\phi}$ is partially entangled state. Moreover, it is assumed that,  one of the two qubits is accelerated uniformly, while the second qubit  is in the  rest. In this  contribution,  we consider  only the second particle is accelerated.

To investigate the behavior of the initial state(1) in non-inertial frame, one has  to describe the
Minkowski coordinates $(t,z)$  by using   Rindler coordinates $(\tau, x)$, which are used   to describe Dirac field in the inertial and  non-inertial frames, respectively. The relations between these
coordinates  are given by\cite{un2},
\begin{equation}\label{trans}
\tau=r~tanh\left(\frac{t}{z}\right), \quad x=\sqrt{t^2-z^2},
\end{equation}
where  $-\infty<\tau<\infty$, $-\infty<x<\infty$  and $r$ is the
acceleration of the moving qubit . The relations (\ref{trans})
describe  two regions in Rindler's spaces: the first region $I $
for $|t|<x$  and the second region $II$ for $x<-|t|$.
 Let us assume that, in the  Minkowski space,  the single mode $k$ of fermions and anti-fermions
is described by the annihilation operators $a_{kU}$  and $b_{kU}$, respectively. However, in terms of Rindler's
operators ( $c^{(I)}_{kR}, d^{(II)}_{-kL}$), the Minkowski
operators may be written as,

\begin{eqnarray}\label{op}
a_{kU}&=&\cos r c^{(I)}_{k,R}-\exp(-i\psi)\sin r d^{(II)}_{-k,L},
\nonumber\\
 b^\dagger_{kU}&=&\exp(i\psi)\sin r
c^{(I)}_{k,R}+\cos r d^{(II)}_{k,L},
\end{eqnarray}
where, $tan~ r=e^{-\pi\omega \frac{c}{a}}$, $0\leq r\leq \pi/$4, $
a$ is the acceleration such that, $0\leq a\leq\infty$, $\omega$ is
the frequency of the traveling qubits, $c$ is the speed of light,
and $\psi$ is an unimportant phase that can be absorbed into the
definition of the operators.
The  operators (\ref{op}) mix a qubit  in region $I$ and an anti qubit  in
region $II$. In  the  computational basis, $|0_k\rangle $ and $|1_k\rangle $,  the Minkowski space of the qubit state are  transformed into  the Rindler space as\cite{Wei}.

\begin{eqnarray}\label{2.4}
|0_k\rangle&=&\cos r |0_k\rangle_{I} |0_k\rangle_{II} + \sin r |1_k\rangle_{I} |1_k\rangle_{II},\quad
 |1_k\rangle= |1_k\rangle_{I} |0_k\rangle_{II}.
\end{eqnarray}

Now, according to our suggested model and   by using Eqs.(1) and Eq.(\ref{op}), the final accelerated state in the presence of the white noise is given by,
\begin{eqnarray}
\hat{{\rho}}^{acc}_w&=&\gamma\cos^2r\ket{00}\bra{00}+\Bigl(\alpha+\gamma\sin^2r\Bigr)\ket{01}\bra{01}+\beta\cos^2r\ket{10}\bra{10}
\nonumber\\
&+&\Bigl(\beta \sin^2 r+\gamma\Bigr)\ket{11}\bra{11}+\epsilon\cos r\Bigl(\ket{01}\bra{10}+\ket{10}\bra{01}\Bigr),
\end{eqnarray}
where
\begin{align*}
\gamma=\frac{1-p}{4},\quad \alpha=\frac{1+p\lparen{3-4x^2}\rparen}{4}, \quad \beta=\frac{1-p\lparen{1-4x^2}\rparen}{4},\quad \epsilon&=px\sqrt{1-x^2}.
\end{align*}

Similarly, in the presences of the color noise, the final  accelerated state is given
by accelerating the second qubit, we get this model:
\begin{align*}
\hat{{\rho}}_c=&{\alpha}_c\ket{01}\bra{01}+{\beta}_c\;\cos^2{r}\ket{10}\bra{10}+{\beta}_c\;\sin^2{r}\ket{11}\bra{11}+{\epsilon}_c \;\cos{r}\lparen{\ket{01}\bra{10}+\ket{10}\bra{01}}\rparen
\end{align*}
where
\begin{equation}
\alpha_c=\frac{1+p\lparen{1-2x^2}\rparen}{2},~\quad {\beta}_c=\frac{1-p\lparen{1-2x^2}\rparen}{2},~\quad {\epsilon}_c=p\;x\sqrt{1-x^2}.
\end{equation}

\section{Entanglement of the accelerated state}
In this section, we use the concurrence \cite{Woottors1,Woottors2} as a measure of  the entanglement of the accelerated system. This measure is defined as,
\begin{equation}
\mathcal{C}=max\{0,\sqrt{\lambda_1}- \sqrt{\lambda_2}-\sqrt{\lambda_3}-\sqrt{\lambda_4}\}
\end{equation}
where $\lambda_1>\lambda_2>\lambda_3>\lambda_4$ and $\lambda_i$ are the eigenvalues of the matrix $\rho_w^{acc}\bar{\rho_{w}^{acc}}$ and $\bar{\rho_{w}^{acc}}=\sigma_y\tau_y{\rho^{*acc}}_w \sigma_y\tau_y$
In this context, we introduce analytical forms of the concurrence for the white, color, and white-color noisy.
\begin{itemize}
\item For the {\it white noise}, the concurrence takes the following form
  \begin{equation}
\mathcal{C}_{w}=max\Big\{{0,-\frac{1}{8}\sqrt{\cos{r}\lparen{w_1-w_2+w_3}\rparen}+\frac{1}{8}\sqrt{\cos{r}\lparen{w_1+w_2+w_3}\rparen}-4\sqrt{w_4}}\Big\}
\end{equation}
where
\begin{align*}
w_1&=\lparen{5+6p-11p^2+4p\lparen{1+31p}\rparen x^2-128p^2x^4}\rparen\cos{r}\\
w_2&=16\sqrt{2}p\;x\;\cos{r}\sqrt{\lparen{1 - x^2}\rparen \lparen{1 - p + 4 p x^2}\rparen\lparen{3 + 5p - 8 p\;x^2-\lparen{1 - p}\rparen \cos{2 r}}}\\
w_3&=\lparen{-1 + p}\rparen \lparen{1 - p + 4 p x^2}\rparen\cos{3 r}\\
w_4&=\cos^2{r}\lparen{1 - p }\rparen\lparen{1 - p  + \lparen{1 + p \lparen{-1 + 4 x^2}\rparen}\rparen \sin^2{r}}\rparen
\end{align*}

\item while in the presence of the {\it color noise}, it takes the form

\begin{equation}
 \mathcal{C}_{c}=max\Big\{
  0,\frac{1}{2}\sqrt{\cos{r}\left(c_1+c_2\right)}-\frac{1}{2}\sqrt{\cos{r}\left(c_1-c_2\right)}
  \Big\},
\end{equation}
with,
\begin{align*}
    c_1&=\cos{r}-p^2\cos{r}\left(1-8x^2+8x^4\right),\\
    c_2&=4p\;x\;\cos{r}\sqrt{\lparen{1-x^2}\rparen\big(1-p^2\lparen{1-2x^2}\rparen^2\big)},
\end{align*}
\item Finally, in the presences of the {\it white-color noises}, the concurrence is given explicitly by
\begin{equation}
\mathcal{C}_{wc}=max\Big\{{0,-\frac{1}{8}\sqrt{a_1-a_2+a_3}+\frac{1}{8}\sqrt{a_1+a_2+a_3}-\frac{1}{2}\sqrt{a_4}}\Big\},
\end{equation}
where
\begin{align*}
a_1&=5+6p-11p^2+8q+8p\;q+3q^2+4p\lparen{1+31p-q}x^2-128p^2x^4\rparen,\\
a_2&=16\sqrt{2}\sqrt{-p^2 x^2 \lparen{-1 + x^2} \lparen{\eta_1+ 4 p x^2} \cos^2{r}\lparen{3 + 5 p + q - 8 px^2 + \eta_2 \cos{2 r}}}),\\
a_3&=\eta_2 \lparen{\eta_1 + 4 p x^2}\rparen\cos{3 r},\\
a_4&=-\eta_2\cos^2{r}\lparen{1 - p - q + \lparen{1 + q + p \lparen{-1 + 4 x^2}\rparen}\rparen \sin^2 r}\rparen,
\end{align*}
\end{itemize}
with $\eta_1=1- p+q$, $\eta_2=-1+p+q$.

The effect of the noise strength, initial state settings on the behavior of the concurrence as a measure of the survival amount of entanglement is described in Figs.(1-4).
\begin{figure}[t!]
	\centering
	\includegraphics[width=0.3\linewidth, height=4cm]{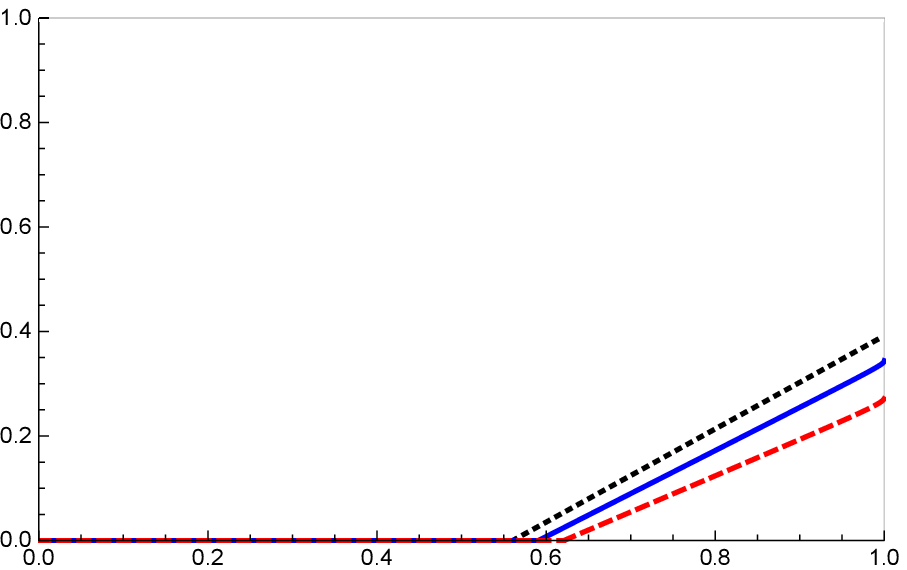}\quad ~
	\includegraphics[width=0.3\linewidth, height=4cm]{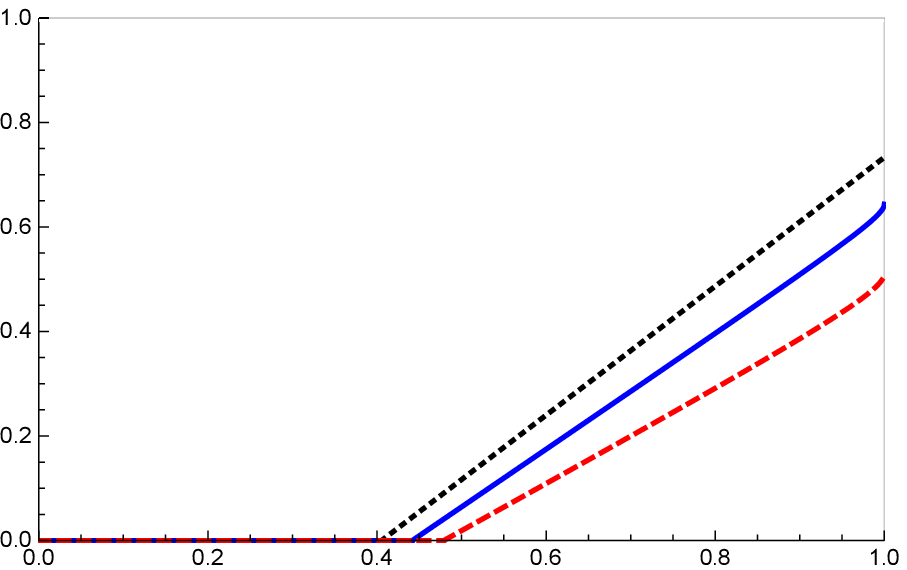}\quad~
\includegraphics[width=0.3\linewidth, height=4cm]{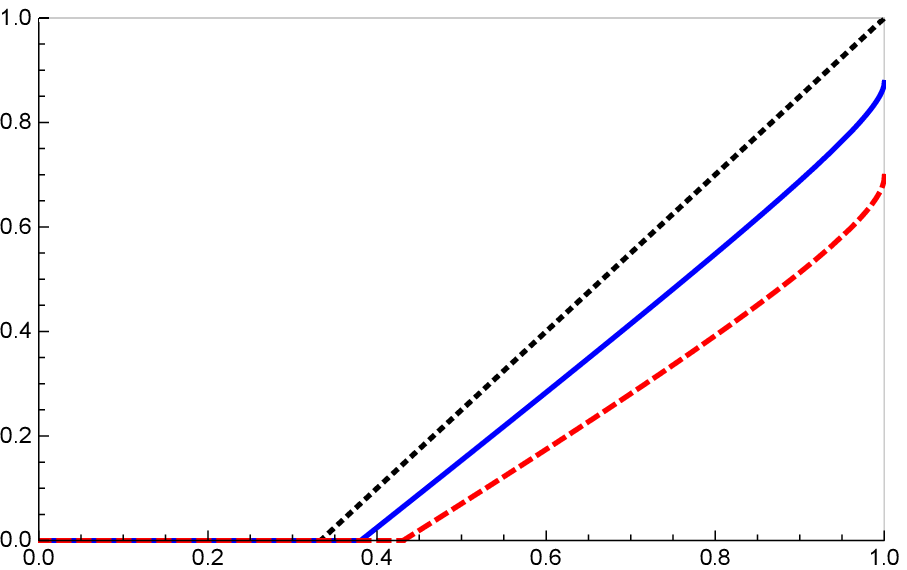}
\put(-60,-10){$p$}
\put(-145,60){$\mathcal{C}_w$}
\put(-230,-10){$p$}
\put(-300,60){$\mathcal{C}_w$}
\put(-370,-10){$p$}
\put(-450,60){$\mathcal{C}_w$}
	\caption{the  concurrence $\mathcal{C}_w$ for the accelerated system in the presence of white noise, where the dot. solid, and dash lines for $r=0,0.5$ and $0.8$, respectively and the system is initially prepared in (a) $x=0.2, (b) x=0.4$ and(c) $x=\frac{1}{\sqrt{2}}$.}
	\end{figure}
In Fig.(1), we investigate the  behavior of the concurrence for different initial state settings. It is clear that,  for small values of $x$, namely the initial state has small amount of entanglement, the  small values of the white noise strength $p$ destroys the of entanglement, even for zero acceleration. However, the entanglement  rebirth again  as the channel parameter $p$ increases, where the re-birthing interval of the channel strength increases if the initial state has a larger degree of entanglement. Moreover, the re-birth entanglement increases monotonically as $p$ increases. The maximum  bounds of entanglement depend on the initial acceleration, where $\mathcal{C}_w=1$ at $r=0$ and $p=1$.

\begin{figure}
	\centering
		\includegraphics[width=0.4\linewidth, height=4cm]{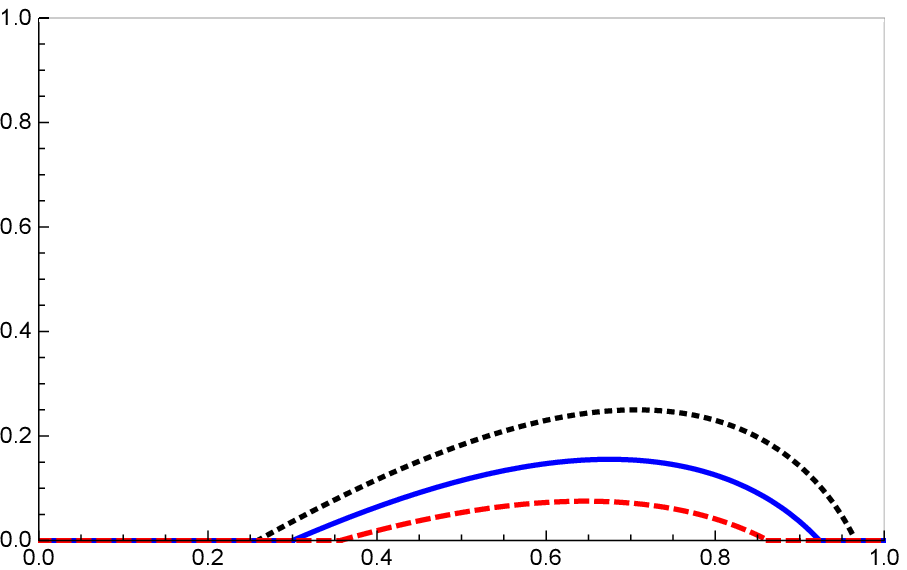}\quad\quad~~~~
\includegraphics[width=0.4\linewidth, height=4cm]{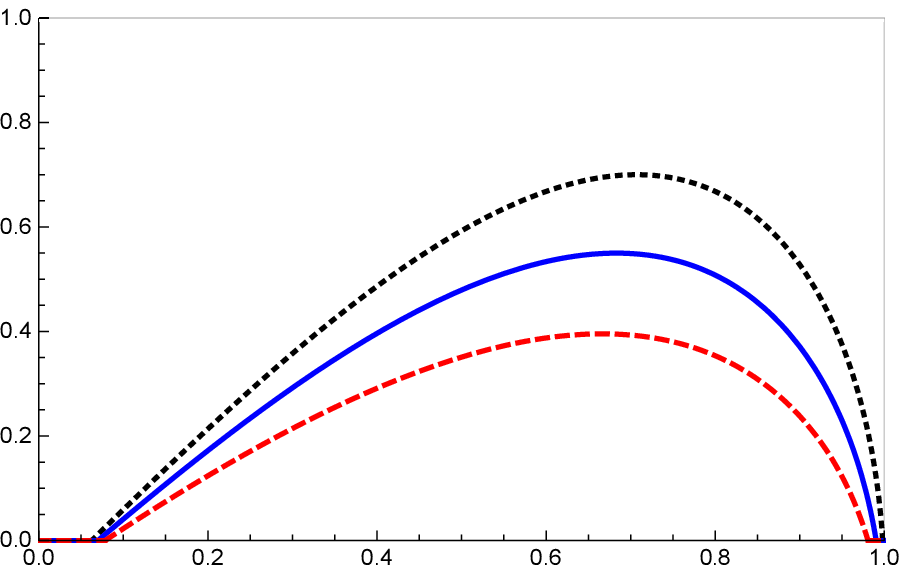}
\put(-30,100){$(b)$}
\put(-250,100){$(a)$}
\put(-90,-10){$x$}
\put(-195,60){$\mathcal{C}_w$}
\put(-300,-10){$x$}
\put(-410,60){$\mathcal{C}_w$}
	\caption{the  concurrence $\mathcal{C}_w$ for  accelerated different initial state settings system in the presence of white noise, where the dot. solid, and dash lines for $r=0.0,0.5$ and $0.8$, respectively, where (a) $ p=0.4$ and(b) $p=0.8$.}
	\end{figure}
Fig.(2) shows the effect of different values of the acceleration in the presence of the white noisy. It is clear that, the entanglement appears at large  at values of $x$  at small channel strength. However as one increases $p$, the entanglement appears at smaller values of the initial state settings. The upper bounds of $\mathcal{C}_w$   depend on the initial acceleration, where the maximum values of  $\mathcal{C}_w$ are depicted at $x=\frac{1}{\sqrt{2}}$ and zero acceleration. However at $x=1$, the initial system  contains only classical correlation and consequently the concurrence $\mathcal{C}_w=0$.

\begin{figure}[t!]
	\centering
	\includegraphics[width=0.3\linewidth, height=4cm]{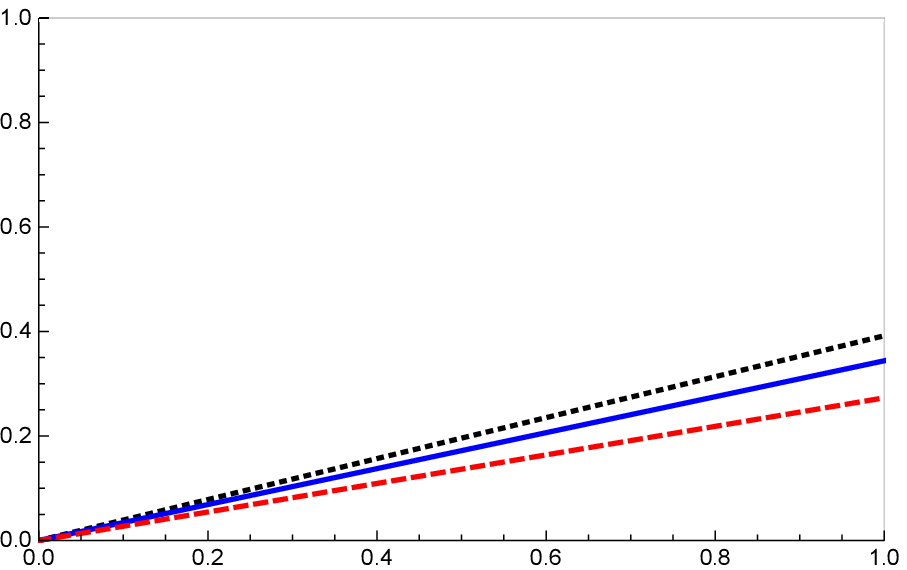}\quad ~
	\includegraphics[width=0.3\linewidth, height=4cm]{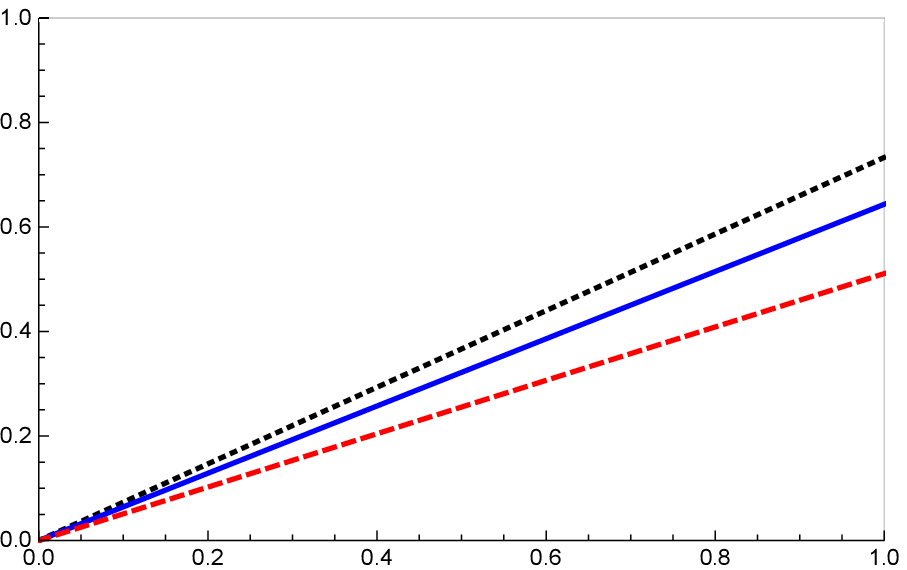}\quad~
\includegraphics[width=0.3\linewidth, height=4cm]{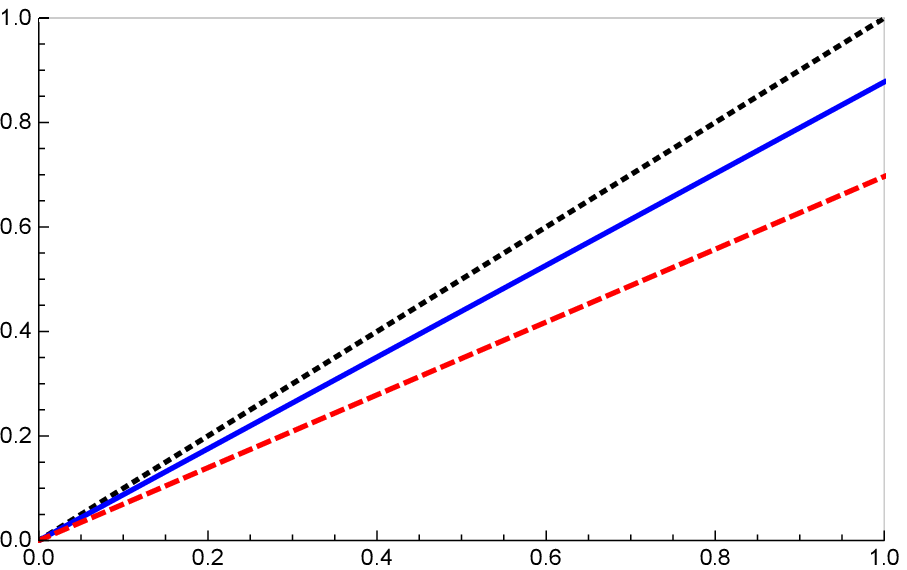}
\put(-60,-10){$q$}
\put(-142,60){$\mathcal{C}_c$}
\put(-220,-10){$q$}
\put(-292,60){$\mathcal{C}_c$}
\put(-370,-10){$q$}
\put(-452,60){$\mathcal{C}_c$}
	\caption{The same as Fig.(1), but in the presence of the color noise.}
	\end{figure}

The behavior of the concurrence, $\mathcal{C}_c$   for the accelerated state in the presence of the {\it color} noise is displayed in Fig.(3), for different initial state settings, where different  acceleration values are considered.  It is clear that, $\mathcal{C}_c$ increases as $q$ increases. However, the increasing rate depends on the initial state settings and the acceleration, where the maximum  bounds are depicted  if the  system is  initially prepared in a MES, namely $x=\frac{1}{\sqrt{2}}$.

From Fig.(1) and Fig.(3) it is clear that, the color noise robust the decoherence due to the acceleration for any $q>0$, while  in the presence  of the white noise, the entanglement appears  at larger values of the channel strength $p$.  Morover, the increasing rate that depicted for the concurrence in the presence of the color noise is much larger than that  shown  for  the white noise.

\begin{figure}
	\centering
		\includegraphics[width=0.4\linewidth, height=4cm]{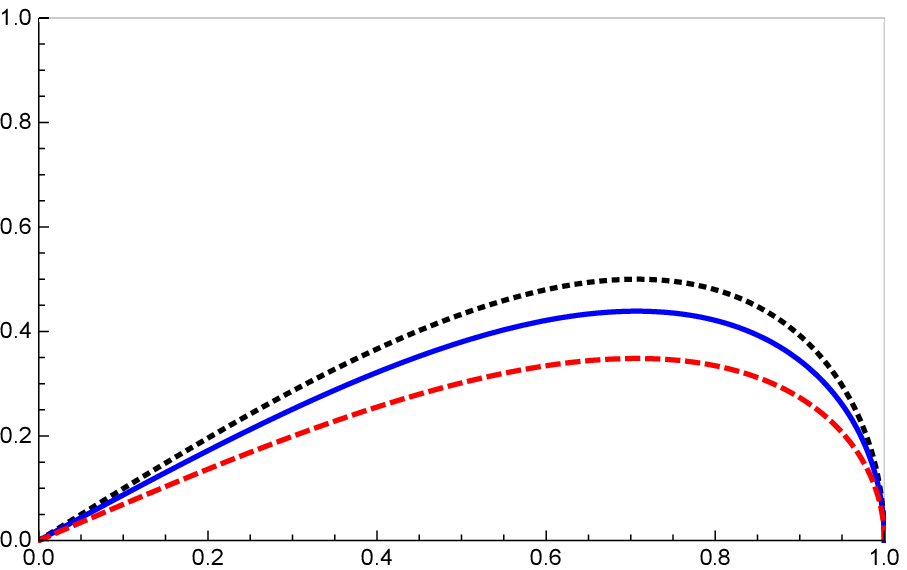}\quad\quad~~~~
\includegraphics[width=0.4\linewidth, height=4cm]{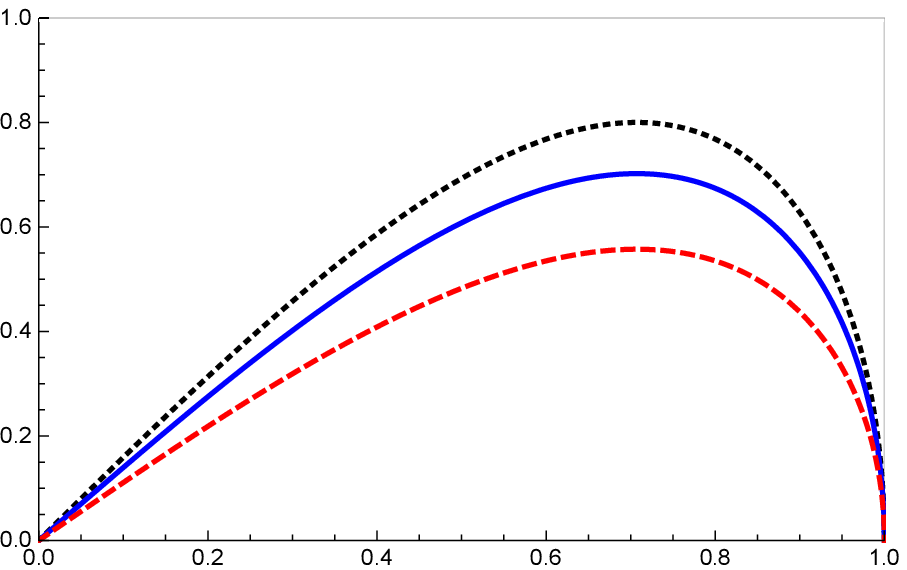}
\put(-30,100){$(b)$}
\put(-250,100){$(a)$}
\put(-90,-10){$x$}
\put(-195,60){$\mathcal{C}_c$}
\put(-300,-10){$x$}
\put(-410,60){$\mathcal{C}_c$}
	\caption{The same as Fig.(2), but in the presence of the color nosy, where (a) $ q=0.4$ and(b) $q=0.8$.}
\end{figure}

In Fig.(4), we show the effect of the color noise on the accelerated system at  different initial accelerations. In general, the behavior is similar to that displayed in Fig.(2), namely in the presence of the white noise. It is clear that,  at  any $x>0$, the entanglement is generated between the two qubits, i.e., $\mathcal{C}_c>0$ and increases gradually to reach its maximum bound and vanishes  suddenly at $x=1$, where the initial system has only classical correlation.  Moreover, as  one increases the color noise strength $q$, the upper bounds  of entanglement  increase, where the maximum values of the concurrence depends on the initial state settings and the initial accelerations.

Form Fig.(2) and (4), we can observe that the color noise enhances the generated entanglement between the two particles even for small values of $x$, which define the purity of the initial state. However, the larger values of the white noise strength improve the generated entanglement.

\begin{figure}[t!]
	\centering
	\includegraphics[width=0.3\linewidth, height=4cm]{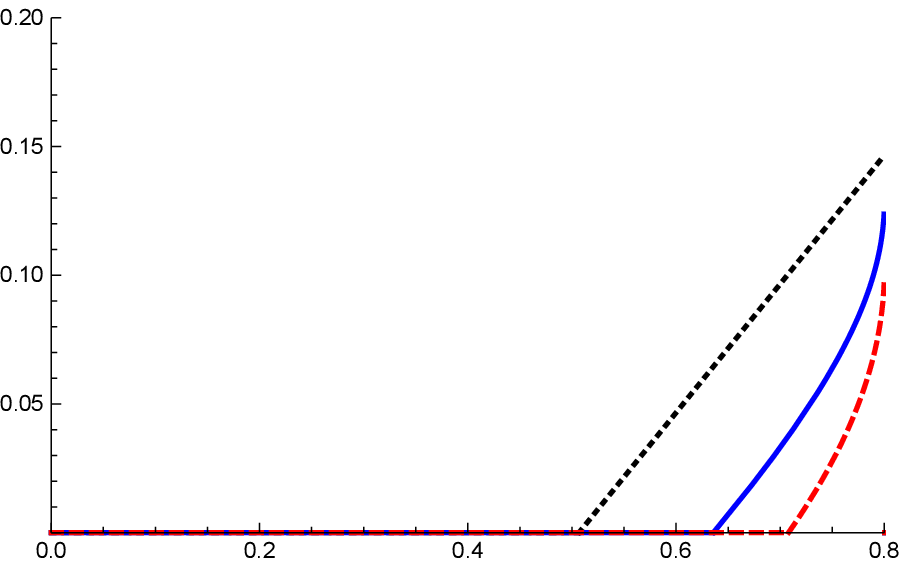}\quad ~
	\includegraphics[width=0.3\linewidth, height=4cm]{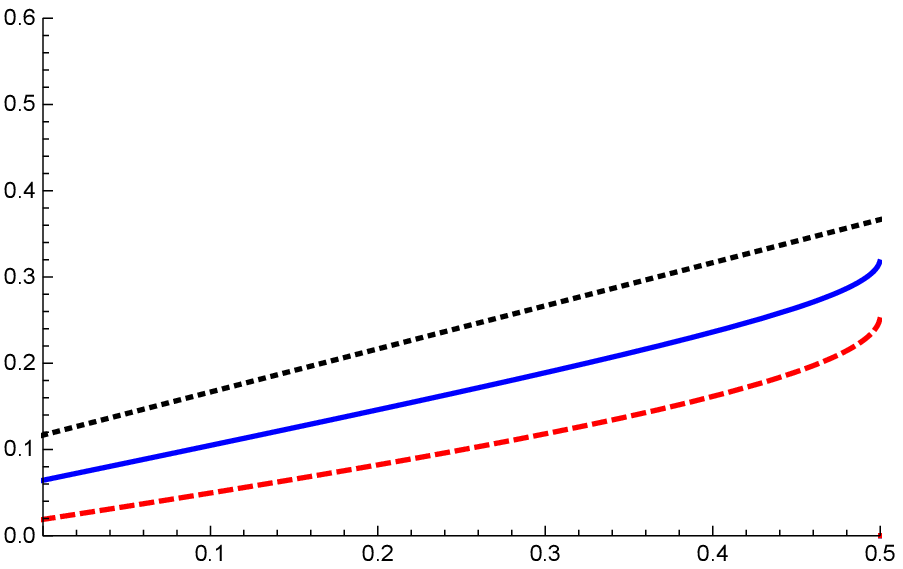}\quad~
\includegraphics[width=0.3\linewidth, height=4cm]{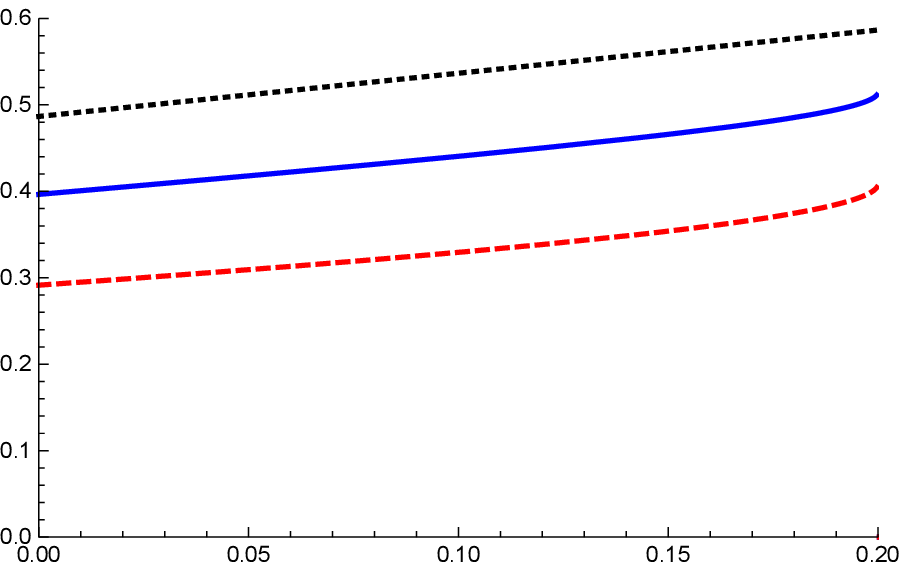}
\put(-62,-10){$q$}
\put(-145,63){$\mathcal{C}_{wc}$}
\put(-415,100){$(a)$}
\put(-250,100){$(b)$}
\put(-230,-10){$q$}
\put(-100,105){$(c)$}
\put(-296,63){$\mathcal{C}_{wc}$}
\put(-370,-10){$q$}
\put(-457,63){$\mathcal{C}_{wc}$}
	\caption{The sane as Fig.(1) but in the presences of white-color noise where the initial system is accelerated by setting $x=0.4$, where (a)
 $p=0.2$, (b)$p=0.5$ and (c) $p=0.8$}
\end{figure}

Fig.(5),  displays  the behavior of the concurrence of the accelerated system when it passes through white-color noise for different values of the  acceleration and the  initial state that  is prepared  at  $x=0.4$.  Fig(5a) shows that, the concurrence is zero at small values of $q$, where we set the white noise strength ($p=0.2)$. As $q$ increases, the entanglement re-births at different values of $q$ depending on the initial accelerations. However, the concurrence $\mathcal{C}_{wc}$ increases gradually as $q$ increases. In Fig.(5b), we increase the value of the white noisy strength, i.e. $p=0.5$. In this case, the entanglement increases at  any value of  $q\in[0,1-p]$, where the maximum bounds depend on the initial acceleration. However, as it is displayed  from  Fig.(5c), the increasing rate of the concurrence is depicted  at large values of the strength  $p=0.8$.

\begin{figure}[t!]
	\centering
		\includegraphics[width=0.4\linewidth, height=4cm]{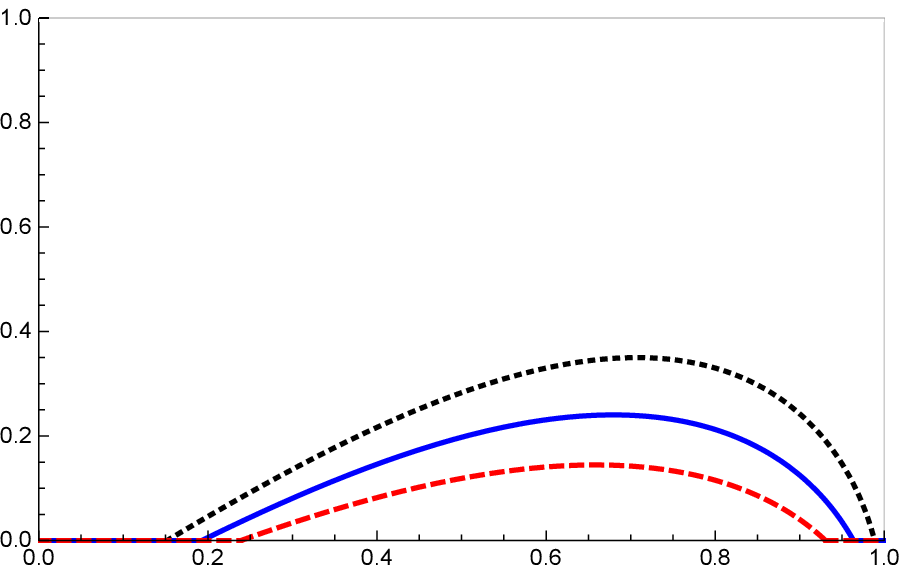}\quad\quad~~~~
\includegraphics[width=0.4\linewidth, height=4cm]{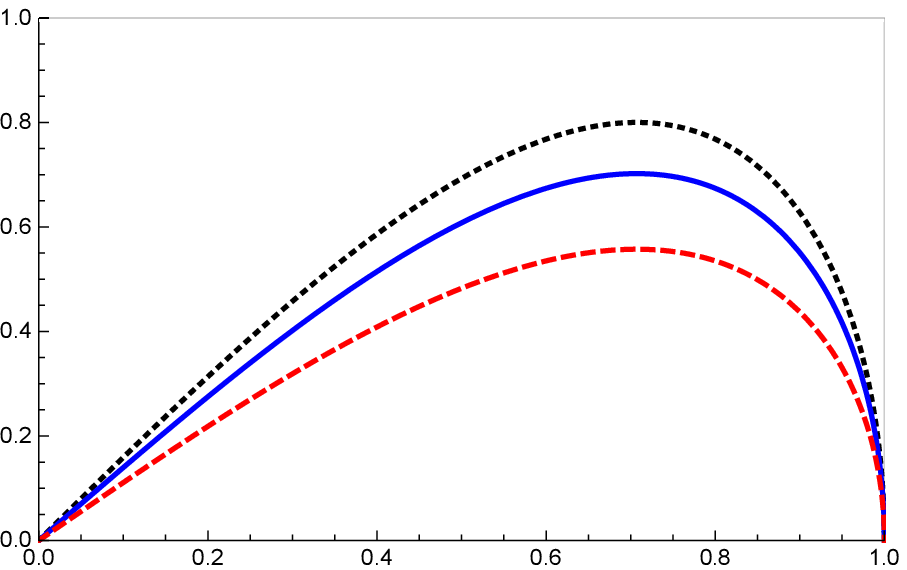}
\put(-30,100){$(b)$}
\put(-250,100){$(a)$}
\put(-90,-10){$x$}
\put(-195,60){$\mathcal{C}_{wc}$}
\put(-300,-10){$x$}
\put(-410,60){$\mathcal{C}_{wc}$}
	\caption{The same as Fig.(2), but in the presence of the white-color  nosy where $q=0.2$ and  (a) $ p=0.5$ and(b) $p=0.8$.}
	\label{fig:4}
\end{figure}

Fig.(6) shows the behavior of $\mathcal{C}_{wc}$ for accelerated system is initially prepared in any state with $x\in[0,~1]$, where we fixed the color  noise  strength parameter ($q=0.2$) and different values of the white color  noise strength $p$ are considered. The general behavior of the concurrence is similar to those shown  Figs.(2) and (4). However, as it is displayed from  Fig.(6a), the non-zero values of the concurrence are displayed at different initial state settings, depending on the initial acceleration of the system. On the other hand, as one increases $p$, the entangled behavior is depicted at any values of $x$ and it is independent of the initial correlation, but the large  degree of entanglement is displayed  for the non-accelerated system.

\begin{figure}
	\centering
		\includegraphics[width=0.4\linewidth, height=4cm]{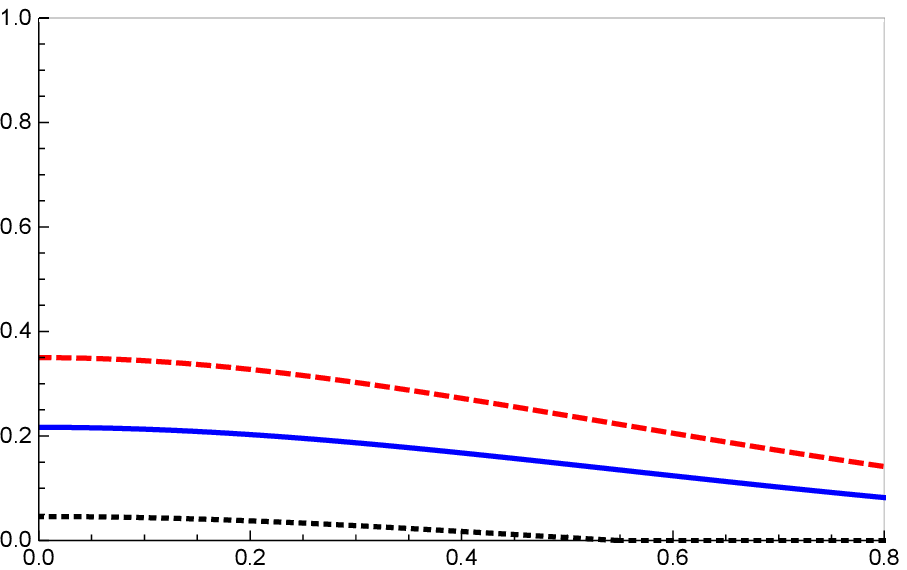}\quad\quad~~~~
\includegraphics[width=0.4\linewidth, height=4cm]{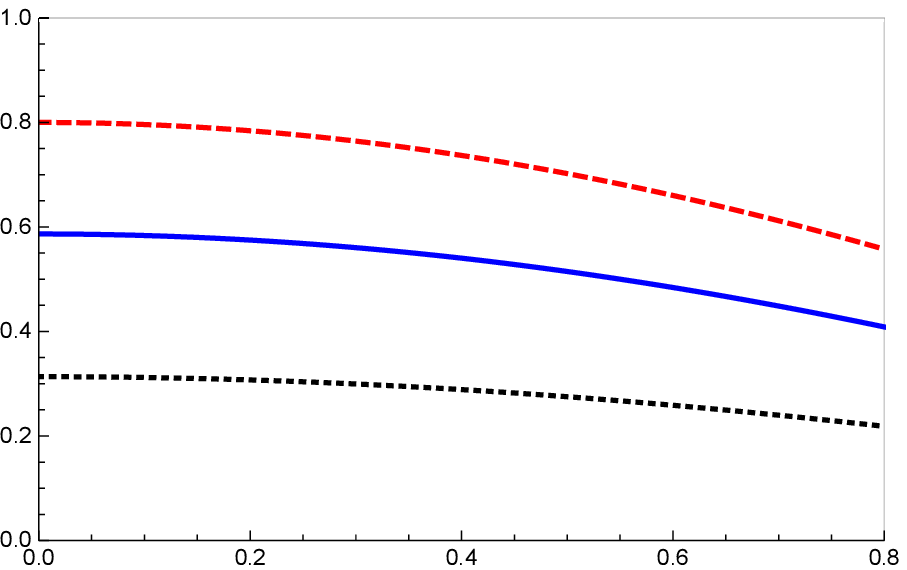}
\put(-30,100){$(b)$}
\put(-250,100){$(a)$}
\put(-90,-10){$r$}
\put(-195,60){$\mathcal{C}_{wc}$}
\put(-300,-10){$r$}
\put(-410,60){$\mathcal{C}_{wc}$}
	\caption{The effect of the acceleration same as Fig.(2), but in the presence of the white-color  nosy where $q=02$ and  (a) $ p=0.5$ and(b) $p=0.8$.}
	\end{figure}

\section{Fisher Information}

The quantum Fisher information   has two different forms, one for the single particle and the second one for a two qubit-system.  For the single qubit it is evaluate by means of the Bloch vector which describe the qubit. However, if the aim is  estimating  a parameter $\beta$ which characterize the single qbit, then $\mathcal{F}_\beta$ is defined as,
\begin{equation}\label{Fisher}
\mathcal{F}_{\beta}=\left\{ \begin{array}{ll}
\frac{\Bigl[\row{s}(\beta)\cdot\frac{\partial{\row{s}(\beta)}}{\partial\beta}\Bigr]^2}{1-\bigl|\row{s}(\beta)\bigr|^2}
+\Bigl(\frac{\partial\row{s}(\beta)}{\partial\beta}\Bigr)^2&,\row{s}(\beta)|<1,\\
\nonumber\\
\Bigl|\frac{\partial\row{s}(\beta)}{\partial\beta}\Bigr|^2 & ~,|\row{s}(\beta)|=1\\
\end{array} \right.
\end{equation}

where $\row{s}(\beta)$ is the Bloch vector of this qubit.

However, for a two qubit system, Fisher information  depends on the eigenvalues and the eigenvectors of the system. Explicitly it takes the form,
\begin{eqnarray}
\mathcal{F}_{\tau }&=&\sum_{i=1}^{n}\frac{1}{\lambda _{i}}\left( \frac{%
\partial \lambda _{i}}{\partial \tau }\right) ^{2}+4\sum_{i=1}^{n}\lambda
_{i}\left( \left\langle \frac{\partial V_{i}}{\partial \tau }\right\vert
\left. \frac{\partial V_{i}}{\partial \tau }\right\rangle -\Big|\left\langle
V_{i}\Big|\frac{\partial V_{i}}{\partial \tau }\right\rangle \Big|%
^{2}\right)
\nonumber\\
 &-&8\sum_{i\neq j}^{n}\frac{\lambda _{i}\lambda _{j}}{\lambda
_{i}+\lambda _{j}}\Big|\left\langle V_{i}\Big|\frac{\partial V_{j}}{\partial
\tau }\right\rangle \Big|^{2},  \label{q2}
\nonumber\\
&=& \Delta_c+\Delta_q-\Delta_p,
\end{eqnarray}
where $\lambda_i, V_i$ are the eigenvalues and the eigenvectors of the system and $\tau$ is the estimated parameter. In what follows, we give explicit froms of the Fisher information when the parameter is estimated either by a single or with the whole system.

\subsection{In the presence of the White Noise}
In this section, we quantify the state parameters $(x,r)$, as well as, the channel parameter, $p$ by using the definition of a single and two qubits. The explicit Fisher information  of these parameters are given analytically as,

\begin{equation}\label{ini}
\mathcal{F}_p=\left\{ \begin{array}{ll}
\frac{2a^2\;\cos^2{r}}{\lparen{1-a\;p}\rparen\lparen{3+a\;p-\lparen{1-a\;p}\rparen\cos{2r}}\rparen}\quad\quad\quad
\lvert{\vec{S}\lparen{p}\rparen}\rvert<1,
\\
\\a^2\;\cos^4{r}\quad\quad\quad\quad\quad\quad\quad\quad\quad\lvert{\vec{S}\lparen{p}\rparen}\rvert=1,
\end{array}\right.
\nonumber\\
\end{equation}

\begin{equation}
\mathcal{F}_x=\left\{\begin{array}{ll}
\frac{32p^2 x^2\cos^2{r}}{\lparen{1-a\;p}\rparen\lparen{3+a\;p-\lparen{1-a\;p\rparen}\cos{2r}}\rparen}\quad\quad\quad{\vec{S}\lparen{x}\rparen}\rvert<1,
\\
\\
16p^2\;x^2\;\cos^4{r}\quad\quad\quad
\quad\quad\quad\quad\lvert{\vec{S}\lparen{x}\rparen}\rvert=1,
\end{array} \right.
\nonumber\\
\end{equation}

\begin{equation}
\mathcal{F}_r=\left\{\begin{array}{ll}
\frac{8\lparen{1-a\;p}\rparen \sin^2{r}}{3+a\;p-\lparen{1-a\;p}\rparen \cos{2r}}\quad\quad\quad\quad\quad\quad\lvert{\vec{S}\lparen{r}\rparen}\rvert<1,
 \\
 \\
 (1-a\;p)^2 \sin^2{2r}\quad\quad\quad\quad\quad\quad\lvert{\vec{S}\lparen{r}\rparen}\rvert=1,
 \end{array}\right.
\end{equation}
where $a=1-2x^2$.

The three parameters $\Delta_i, i=c,q$ and $p$ which describe the Fisher information for the two-qubits Eq.(10) are given by

\begin{itemize}
\item Fisher information with respect to the parameter $p$, $\mathcal{F}_P$

\begin{eqnarray}
 \Delta^pc&=&\frac{{{\gamma}\;'}^2\cos^2{r}}{\gamma}+\frac{\lparen{{\kappa}\;'_1-{{\kappa}\;'_2}}\rparen^2}{16\lparen{{\kappa}_1-{\kappa}_2}\rparen}+\frac{\lparen{{\kappa}\;'_1+{{\kappa}\;'_2}}\rparen^2}{16\lparen{{\kappa}_1+{\kappa}_2}\rparen}+\frac{\lparen{{\gamma}\;'+{\beta}\;'\sin^2{r}}\rparen^2}{{\gamma}+{\beta}\sin^2{r}},
\nonumber\\
 \Delta^p_p&=&\frac{1}{4}\bigg[{\frac{\left({{\kappa}\;_1-{\kappa}\;_2}\right)\;{\mu}_1'^2}{\lparen{{\mu}_1^2+1}\rparen^2}+\frac{\lparen{{\kappa}\;_1+{\kappa}\;_2}\rparen\;{\mu}_2'^2}{\lparen{{\mu}_2^2+1}\rparen^2}}\bigg],
 \nonumber\\
 \Delta^p_m&=&8\;\frac{{\lambda}_2{\lambda}_3}{{\lambda}_2+{\lambda}_3}\bigg[{\frac{\left({{\mu}_1-{\mu}_2}\right)^2}{\left({{\mu}_1^2+1}\right)\left({{\mu}_2^2+1}\right)}\bigg({\frac{{\mu}_1'^2}{\left({{\mu}_1^2+1}\right)^2}+\frac{{\mu}_2'^2}{\left({{\mu}_2^2+1}\right)^2}}\bigg)}\bigg],
 \end{eqnarray}
where
 \begin{eqnarray}
 {\kappa}_i'=\kappa_i p\;\;\;\;\;i=1,2,3,~\quad
 {\mu}_i'=\mu_i p\;\;\;\;\;i=1,2
  \end{eqnarray}

\item Fisher information w.r. t to the  parameter $x$, $\mathcal{F}_x$ are given by,
\begin{eqnarray}
 \Delta^x_c&=&\frac{\lparen{{\kappa}\;'_1-{{\kappa}\;'_2}}\rparen^2}{16\lparen{{\kappa}_1-{\kappa}_2}\rparen}+\frac{\lparen{{\kappa}\;'_1+{{\kappa}\;'_2}}\rparen^2}{16\lparen{{\kappa}_1+{\kappa}_2}\rparen}+\frac{\lparen{{\beta}\;'\sin^2{r}}\rparen^2}{{\gamma}+{\beta}\sin^2{r}},
\nonumber\\
 \Delta^x_p&=&\frac{1}{4}\bigg[{\frac{\left({{\kappa}\;_1-{\kappa}\;_2}\right)\;{\mu}_1'^2}{\lparen{{\mu}_1^2+1}\rparen^2}+\frac{\lparen{{\kappa}\;_1+{\kappa}\;_2}\rparen\;{\mu}_2'^2}{\lparen{{\mu}_2^2+1}\rparen^2}}\bigg],
 \nonumber\\
 \Delta^x_m&=&8\;\frac{{\lambda}_2{\lambda}_3}{{\lambda}_2+{\lambda}_3}\bigg[{\frac{\left({{\mu}_1-{\mu}_2}\right)^2}{\left({{\mu}_1^2+1}\right)\left({{\mu}_2^2+1}\right)}\bigg({\frac{{\mu}_1'^2}{\left({{\mu}_1^2+1}\right)^2}+\frac{{\mu}_2'^2}{\left({{\mu}_2^2+1}\right)^2}}\bigg)}\bigg],
 \end{eqnarray}
 where in this case,
 \begin{eqnarray}
 {\kappa}_i'=\kappa_i x\;\;\;\;\;i=1,2,3, ~\quad {\mu}_i'=\mu_i x\;\;\;\;\;i=1,2
  \end{eqnarray}

 \item For the  parameter $r$, $\mathcal{F}_r$ :
\begin{eqnarray}
 \Delta^r_c&=&\frac{\sin{2r}}{\cos^2{r}}+\frac{\lparen{{\kappa}\;'_1-{{\kappa}\;'_2}}\rparen^2}{16\lparen{{\kappa}_1-{\kappa}_2}\rparen}+\frac{\lparen{{\kappa}\;'_1+{{\kappa}\;'_2}}\rparen^2}{16\lparen{{\kappa}_1+{\kappa}_2}\rparen}+\frac{\lparen{{\beta}\;'\sin^{2r}}\rparen^2}{{\gamma}+{\beta}\sin^2{r}},
\nonumber\\
\Delta^r_p&=&\frac{1}{4}\bigg[{\frac{\left({{\kappa}_1-{\kappa}_2}\right)\;{\mu}_1'^2}{\lparen{{\mu}_1^2+1}\rparen^2}+\frac{\lparen{{\kappa}_1+{\kappa}_2}\rparen\;{\mu}_2'^2}{\lparen{{\mu}_2^2+1}\rparen^2}}\bigg],
\nonumber\\
 \Delta^r_m&=&8\;\frac{{\lambda}_2{\lambda}_3}{{\lambda}_2+{\lambda}_3}\bigg[{\frac{\left({{\mu}_1-{\mu}_2}\right)^2}{\left({{\mu}_1^2+1}\right)\left({{\mu}_2^2+1}\right)}\bigg({\frac{{\mu}_1'^2}{\left({{\mu}_1^2+1}\right)^2}+\frac{{\mu}_2'^2}{\left({{\mu}_2^2+1}\right)^2}}\bigg)}\bigg],
 \end{eqnarray}
 where
 \begin{eqnarray}
 {\kappa}_i'&=&\kappa_i r, ~i=1,2,3,\quad {\mu}_i'=\mu_i r \;\;\;\;\;i=1,2
  \end{eqnarray}
where
\begin{align*}
{\kappa}_1&=4+4p-4p\;x^2+4p\;x^2\;\cos{2r},
\\{\kappa}_2&=\sqrt{b_1-b_2+b_3},
\\{\kappa}_3&=2+6-12p\;x^2-2\;\cos{2r}+2p\;\cos{2r}-4p\;x^2\;\cos{2r},\\
b_1&=6+20p+38p^2-40p\;x^2-24p^2x^2+24p^2x^4,\\
b_2&=8\;\cos{2r}\lparen{1+p\lparen{2-4x^2}\rparen+p^2\lparen{-3-4x^2+4x^4}\rparen}\rparen,\\
b_3&=2\;\cos{4r}\lparen{-1+p-2p\;x^2}\rparen^2,\\
{\mu}_1&=\frac{\sec{r}\;}{16{\epsilon}}\lparen{{\kappa}_3-{\kappa}_2}\rparen,
\\{\mu}_2&=\frac{\sec{r}}{16{\epsilon}}\lparen{{\kappa}_3+{\kappa}_2}\rparen.
 \end{align*}
\end{itemize}

\begin{figure}[t!]
	\centering
	\includegraphics[width=0.4\linewidth, height=4cm]{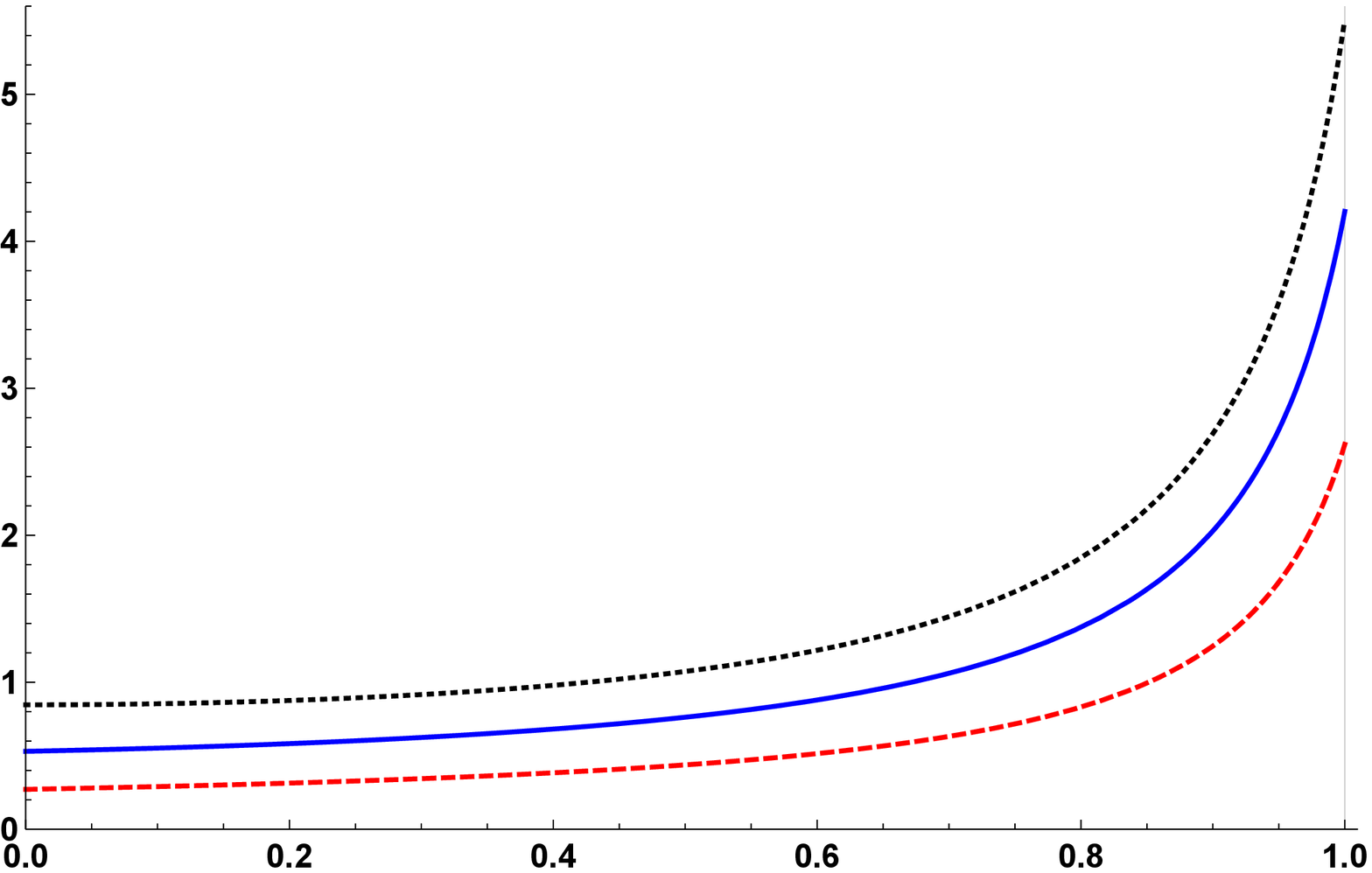}~~~\quad\quad\quad
	\includegraphics[width=0.4\linewidth, height=4cm]{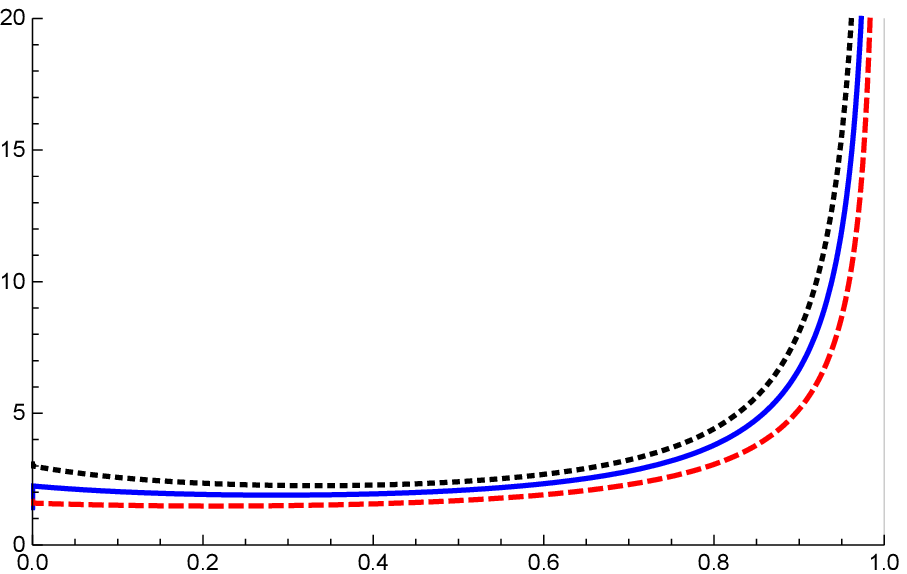}
\put(-90,-10){$p$}
\put(-200,70){$\mathcal{F}_p$}
\put(-300,-10){$p$}
\put(-420,70){$\mathcal{F}_p$}
	\caption{Estimating the Fisher information $\mathcal{F}_p$ for the accelerated system in the presence of white noise where the dot, solid, and dash lines for $r=0,0.5$ and $0.8$, respectively and the system is initially prepared with $x=0.2$  by using (a) Single -qubit and (b) Two-qubits.}
	\end{figure}

\begin{figure}[t!]
	\centering
	\includegraphics[width=0.4\linewidth, height=4cm]{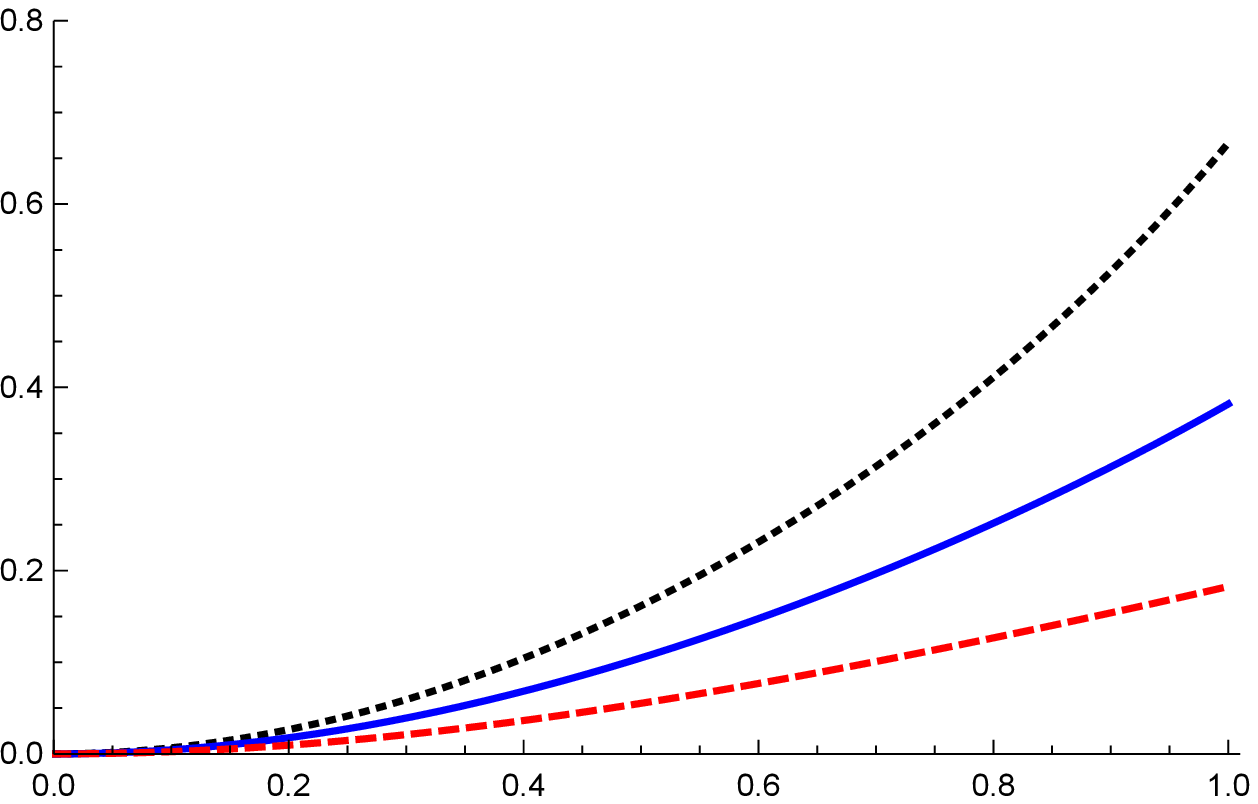}~~~\quad\quad\quad
	\includegraphics[width=0.4\linewidth, height=4cm]{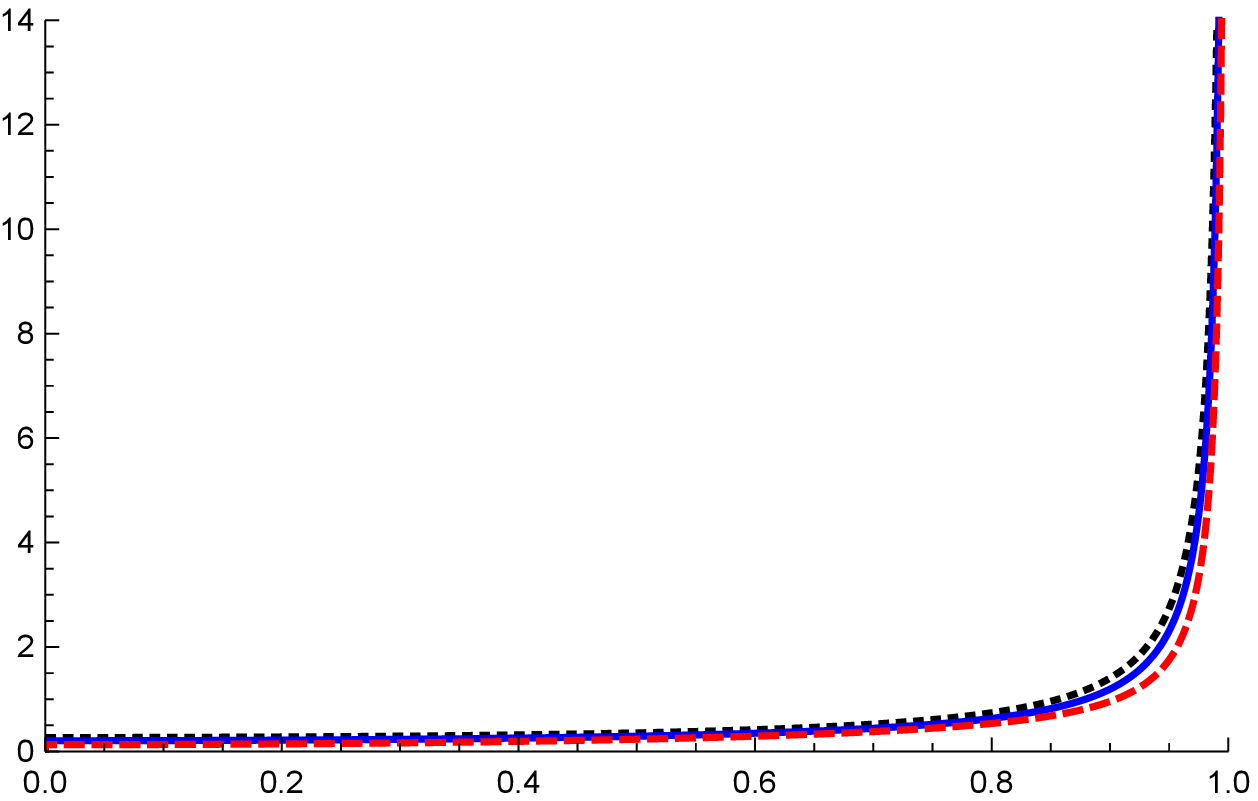}\quad~
\put(-90,-10){$x$}
\put(-200,70){$\mathcal{F}_x$}
\put(-300,-10){$x$}
\put(-420,70){$\mathcal{F}_x$}
	\caption{Estimating the Fisher information $\mathcal{F}_x$ for the accelerated system in the presence of white noise where the dot, solid, and dash lines for $r=0,0.5$ and $0.8$, respectively and the system is initially prepared with $p=0.2$  by using the form of a (a) single -qubit and (b) two-qubits.}
	\end{figure}

\begin{figure}
	\centering
	\includegraphics[width=0.4\linewidth, height=4cm]{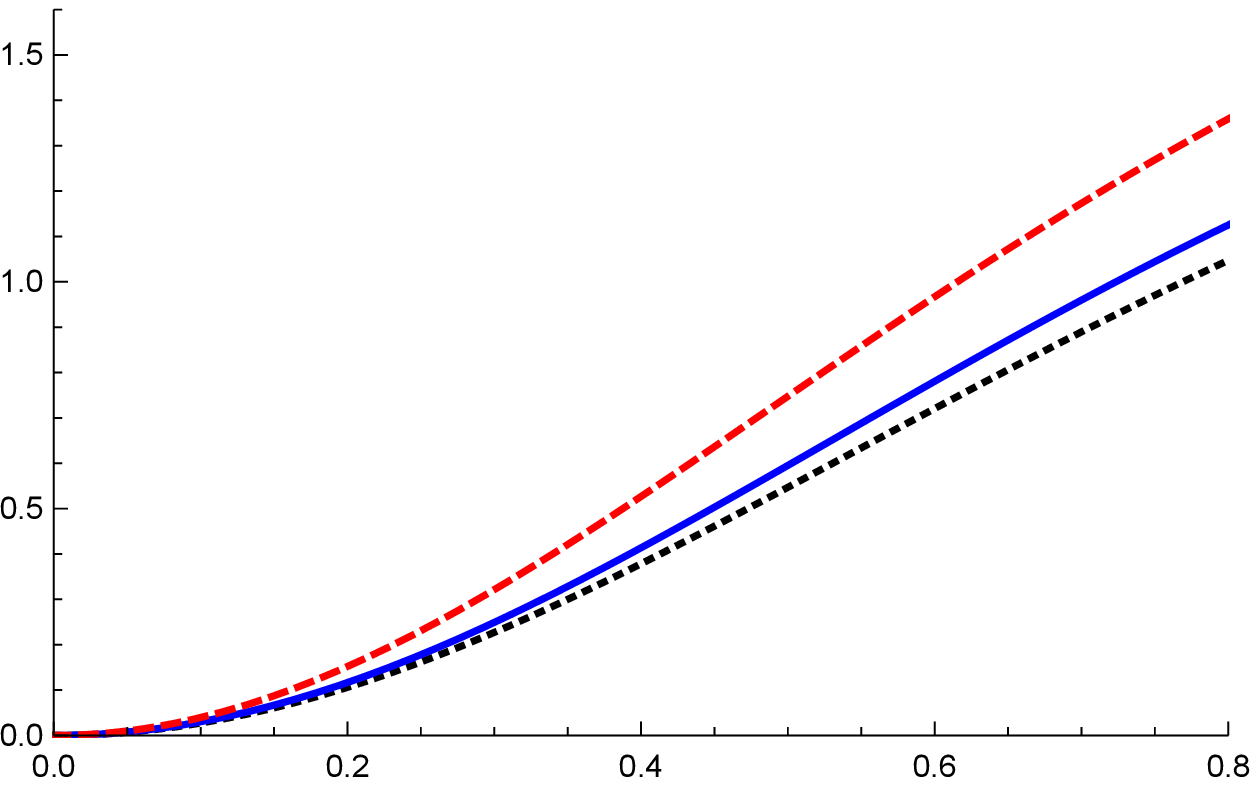}\quad\quad\quad~~
	\includegraphics[width=0.4\linewidth, height=4cm]{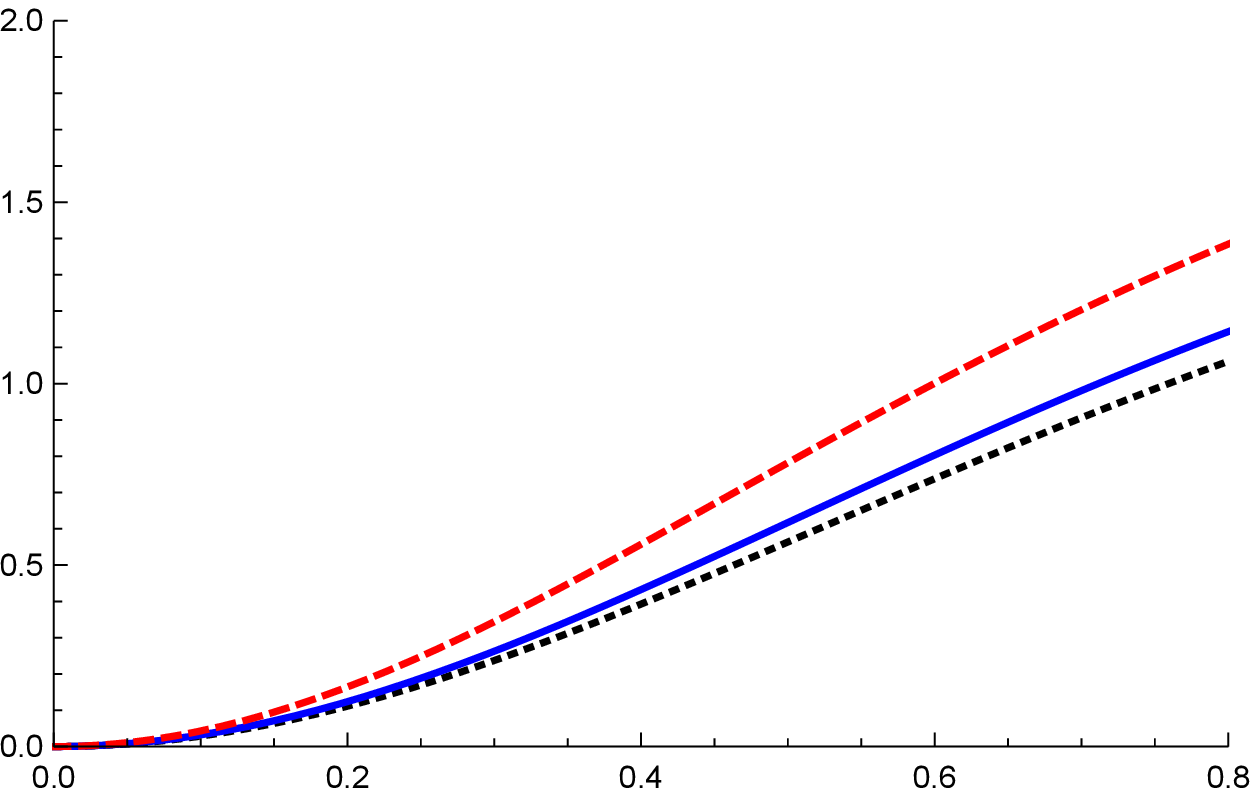}
\put(-90,-10){$r$}
\put(-200,70){$\mathcal{F}_r$}
\put(-300,-10){$r$}
\put(-420,70){$\mathcal{F}_r$}
	\caption{Estimating the Fisher information $\mathcal{F}_r$ for the accelerated system in the presence of white noise where the dot, solid, and dash lines for $x=0.2,  0.4$ and $\frac{1}{\sqrt{2}}$, respectively and  $p=0.2$ ,by using the form of the  (a) single -qubit and (b) the two-qubits.}
	\end{figure}

 In Fig.(8), we quantity the Fisher information with respect to the noisy channel parameter $\mathcal{F}_p$ by using a single and two-qubit formals, where different initial values of the acceleration are considered. The general behavior of $\mathcal{F}_p$  that displayed in Fig.(8a)  and (8b) is similar. However, for $\mathcal{F}_p$ that depicted by using the two-qubit formal is much larger than that displayed by using the single qubit. Moreover, the maximum values of $\mathcal{F}_p$ are displayed at larger values of $p$ and small values of $r$.

Fisher information with respect to the initial state settings $\mathcal{F}_x$ is displayed in Fig.(9). The behavior shows that, $\mathcal{F}_x$ increases as $x$ increases and the maximum values of the Fisher information are  displayed at $x=1$, namely $\ket{\phi}$  contains only classical information.

Finally in Fig.(10), the behavior of $\mathcal{F}_r$  is displayed for different initial state settings. From Figs.(10a) and (10b), it is clear that the behavior is similar for the two used forms. However, $\mathcal{F}_r$, that depicted by using the two -qubit form is a little bit larger. The Fisher information $\mathcal{F}_r$ increases gradually  as the acceleration increases. The upper bounds depends on the initial state settings.

From Figs.(8-10) it is clear that, it is possible to estimate the parameters that describe the accelerated state either by using a single or two-qubit formals. For all cases, it is shown that,  by using the two-qubit formal, the estimation degree of these parameters is larger than that displayed by using a single-qubit form

In Fig.(1),The behavior of the three Fisher information quantities, $\mathcal{F}_p, \mathcal{F}_x$ and $\mathcal{F}_r$ are discussed in the presence of the color noise. Our finding shows that, by using the single qubit- form the behavior of the three quantities are is similar to that predicted for the white noise.
Also, by using the two-qubit form, the same behavior is predicted, namely, they increase gradually as the noises  parameters are increases. However, in the presence of the color noise, the effect of the acceleration  on the estimation degree of these  the channel strength parameter $p$ and the initial state setting parameter $x$, appears only at larger initial values  of these parameters. Moreover, for  $\mathcal{F}_r$ the behavior is completely different, where it is independent from the initial acceleration, but depends on the initial parameters that describe the accelerated state

\begin{figure}[t!]
	\centering
		\includegraphics[width=0.3\linewidth, height=4cm]{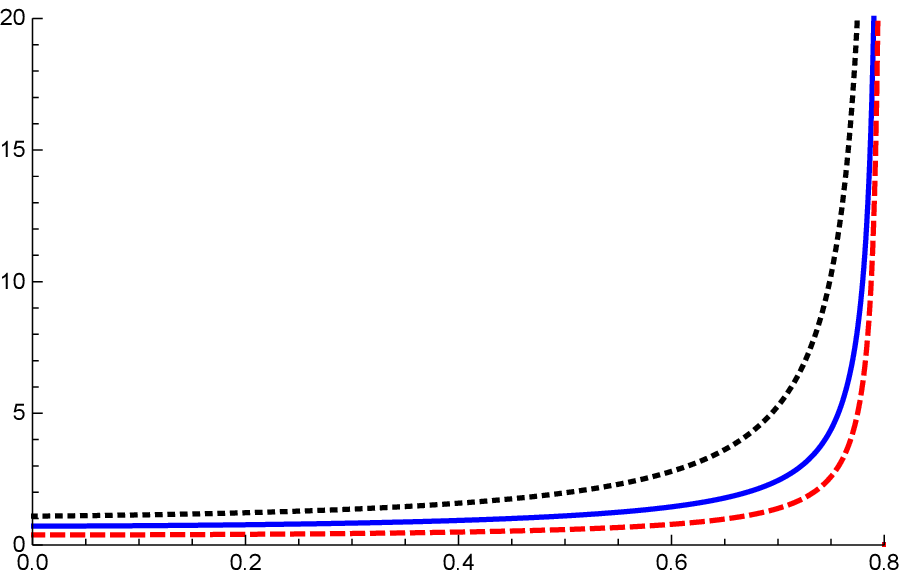}\quad ~
	\includegraphics[width=0.3\linewidth, height=4cm]{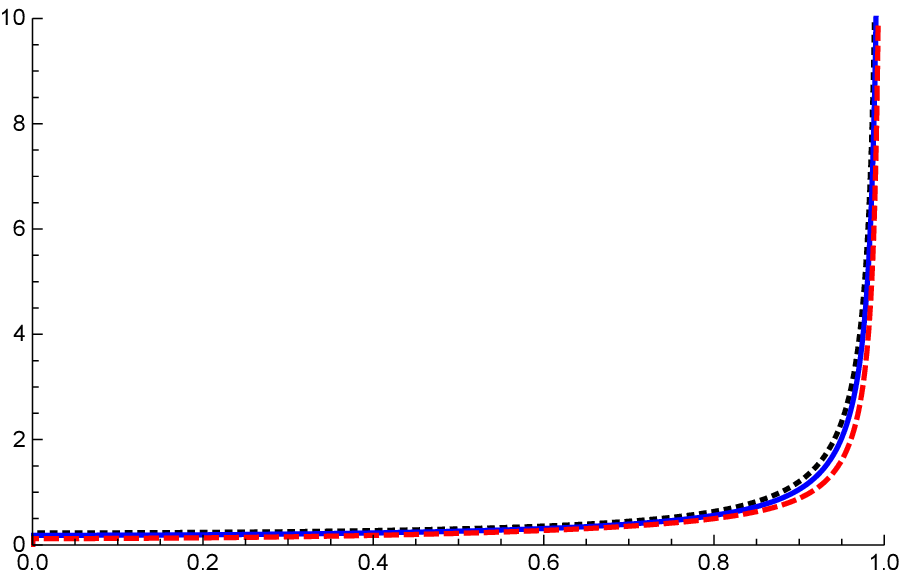}\quad~
\includegraphics[width=0.3\linewidth, height=4cm]{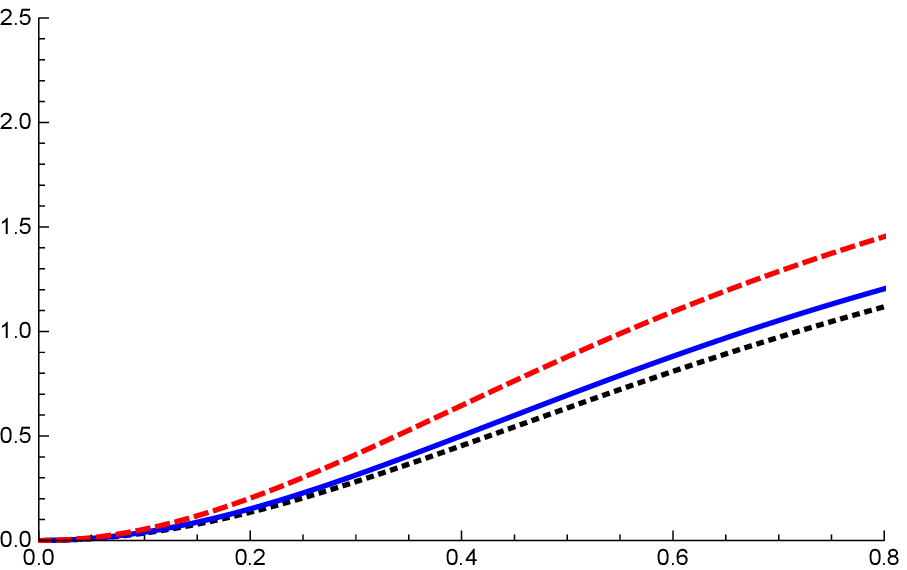}
\put(-60,-10){$p$}
\put(-145,60){$\mathcal{F}_p$}
\put(-230,-10){$p$}
\put(-300,60){$\mathcal{F}_x$}
\put(-370,-10){$p$}
\put(-450,60){$\mathcal{F}_r$}
	\caption{ The same as Fig.(8), but for the color noise.}
	\end{figure}

\section{Conclusion}
In this manuscript, a system consists of two qubit is initially prepared in different initial state settings. It could be maximum, partial entangle state or separable state.  We consider, only one system is accelerated and during this process the two qubits are subject to different types of noisy, either white or color noise. The decoherence will take place for this system due to the acceleration and the noisy channels. The aim of this contribution   is to discusses two tasks; the  first is quantifying the amount of the survival amount of entanglement via the concurrence, while the second is  estimating the initial parameters that describe the accelerated system by using the Fisher information. To achieve the estimation process, we consider two different forms, one depends on a single qubit and the second on the two-qubits.
We  introduced  the solution of this system analytically, where we give   analytical forms of the concurrence in the presences of the white and the color noises. Also, we give a closed form  of the Fisher information, if we use a single or two-qubits form. Moreover, we show  numerically, the effect of the channel strength, acceleration and the initial state settings parameter on these two physically quantities, entanglement and the Fisher information.

The numerical results of the concurrences shows that, the entanglement increases as the channel strengths of the white noise  increases. However, for the white noise the entanglement between the two qubits is generated at different intervals of the channel strength. These intervals depend on the initial state settings, where at  small values of the initial state settings, the entanglement is generated at larger values of the channel strength. Also, for the non-accelerated system the entanglement is generated at smaller values of the channel strength. The maximum values of entanglement is achieved if the system is prepared initially in a  maximum entangled state and smaller values of the channel strength.
 A similar effect of the color noise is depicted for the entanglement, but the entanglement  is generated and increases suddenly  for any value of the  channel strength. Also, it is generated and increases gradually  at any values of the initial state setting's parameter. The maximum bounds  of entanglement are displayed if the system is initially prepared in maximum entangled state, large values of the channel strength, and small values of the acceleration.

The effect of the presences of both noises on the accelerated system is discussed, where  it is shown that both strengths play an important role to increases the entanglement, where we assume that, the accelerated system is initially prepared in  a partially entangled state.  We show that, if one  increases the white noise strength, the entanglement is generated at any values of the color noise strength. However, at small values of the white noise strength, the entanglement is generated at large values of the color noisy' strength.

We estimate the initial parameters that describe the accelerated state, as well as,  the   strength  of the noisy channels, by  evaluation the corresponding Fisher information. Two forms are used to quantify the Fisher information, either by using the single/ two qubits form. The general behavior that predicted for all the estimated parameters is similar  for both forms. However, for all types of noisy channels, the upper bounds of the estimation degree of any parameter that displayed by using the single qubit form is much smaller than  that displayed if the two-qubit's form is used. Moreover,  the variation of the upper bounds of the estimation degree  that depicted by using a single qubit form at different accelerations are much larger than those displayed for  two-qubit's form.

\end{document}